\documentclass[aps,prx,amsmath,amssymb,twocolumn,twoside,showpacs,superscriptaddress,longbibliography]{revtex4-2}
\usepackage[pdftex]{graphicx}
\usepackage[colorlinks=true,citecolor=blue]{hyperref}
\usepackage{graphicx,textcomp,dcolumn,hyperref,upgreek}

\usepackage{physics}

\usepackage{url}

\usepackage{lipsum}

\usepackage{dsfont}

\usepackage{blindtext}

\begin{document}

\title{
Observation of correlations beyond the single-particle picture in fermionic counting statistics
}

\author{Johann~Z{\"o}llner}
\email{johann.zoellner@uni-due.de}
\affiliation{Faculty of Physics and CENIDE, University of Duisburg-Essen, 47057 Duisburg, Germany}

\author{Hendrik Mannel}
\affiliation{Faculty of Physics and CENIDE, University of Duisburg-Essen, 47057 Duisburg, Germany}
\affiliation{Department for Quantum Technology, University of Münster, 48149 Münster, Germany}

\author{Daniel Oppers}
\affiliation{Faculty of Physics and CENIDE, University of Duisburg-Essen, 47057 Duisburg, Germany}

\author{Martin Geller}
\affiliation{Faculty of Physics and CENIDE, University of Duisburg-Essen, 47057 Duisburg, Germany}

\author{Axel Lorke}
\affiliation{Faculty of Physics and CENIDE, University of Duisburg-Essen, 47057 Duisburg, Germany}

\author{J{\"u}rgen~K{\"o}nig}
\affiliation{Faculty of Physics and CENIDE, University of Duisburg-Essen, 47057 Duisburg, Germany}
               
\date{\today}

\begin{abstract}
    The analysis of charge fluctuations in a mesoscopic device can shed light on the fundamental question of whether or not a single-particle
    picture of electron transport is appropriate, even in the presence of Coulomb interaction.
    It has been shown on general grounds that mapping the interacting system onto a fermionic single-particle description is subject to strict constraints on the charge-transfer statistics. These constraints can be formulated in terms of a sign rule for the factorial cumulants of the full counting statistics.
    We report here on the experimental observation of a sign-rule violation in a most elementary model system, i.e. a single-level quantum dot, and provide theoretical modeling that can quantitatively reproduce the experimental data.  
    This reveals the presence of correlations introduced by electron-electron interaction.
\end{abstract}

\maketitle

\section{Introduction}

Electric currents in solid-state devices are carried by electrons.
In addition to being passively transported, the charged electrons often play an active role in establishing, through mutual Coulomb interaction, a complicated many-body system.
Despite the potential complexity of such systems, transport through them can often be described within a scattering-matrix approach, that intrinsically includes correlations due to the Pauli exclusion principle as well as interactions of the electrons with impurities or the potential defining the device.
It is remarkable that the underlying single-particle picture then incorporates the two-particle interactions through modified scattering amplitudes.
A prominent example of strongly-correlated many-body physics is the Kondo effect in metals due to magnetic impurities.
At low temperatures, the spin fluctuations are screened such that the impurities merely appear as potential scatterers for single electrons but with a sharp resonance near the Fermi edge \cite{hewsonKondo1993}.
Similarly, the Stoner model, which treats the collective magnetic order within a single-particle picture simply by spin-split bands, is often sufficient to properly model transport in ferromagnets \cite{marderCondensed2010}.

Whether or not a system can be described within a single-particle picture has been answered by the Abanov-Ivanov theorem \cite{abanovFactorization2009}.
It states that the charge-transfer statistics of \textit{non-interacting} fermions through \textit{any} two-terminal device can be decomposed into independent \textit{single-particle} events with, in general, different transfer probabilities $p_j \in [0,1]$ for each charge-transfer channel $j=1,2,...$ .
This theorem, whose proof is based on the determinant formula introduced by Levitov and Lesovik \cite{levitovCharge1993,ivanovStatistics1993,ivanovCoherent1997}, holds for any temperature, voltage or scattering matrix and generalizes earlier work on fundamental concepts of transport statistics of tunneling electrons \cite{vanevicElementary2007,hasslerWavepacket2008,vanevicElementary2008, abanovAllowed2008}. 
In the presence of electron-electron interaction, the Abanov-Ivanov theorem is no longer applicable.
Nevertheless, the charge-transfer statistics may still, in many cases, be described within a single-particle picture, but with renormalized transfer probabilities $p_j$.
A proof of interaction-induced correlations is only given once the decomposition into independent single-particle events is no longer possible.
This is somewhat analogous to prove quantum entanglement by violation of the Bell inequalities in correlation measurements.

\section{Theory}

According to the Abanov-Ivanov theorem, unidirectional transport of non-interacting electrons through an arbitrary scattering region can be decomposed into single charge-transfer events.
This means that the probability $P_N$ (also referred to as the \textit{full counting statistics}) that $N$ charges have been transferred between two contacts within a given time interval can be written in the form
\begin{align}
    P_N =\sum_{A_N\subset \mathbb{N}}\prod_{j\in A_N}p_j\prod_{k \in \mathbb{N} \backslash A_N}(1-p_k) \, ,
    \label{eq:generalized binomial}
\end{align}
where the sum goes over all natural-number subsets $A_N$ with $N$ elements, and $\mathbb{N} \backslash A_N$ is the complement of $A_N$.
Within a scattering picture, $j$ is interpreted as a channel index and $p_j\in [0,1]$ as the transmission probability through this channel.
Equation~\eqref{eq:generalized binomial} has been dubbed \textit{Poisson's binomial distribution} \cite{wangNumber1993} or \textit{generalized binomial distribution} \cite{hasslerWavepacket2008,abanovFactorization2009}. It generalizes the binomial distribution $P_N^{\text{\,binomial}}= \binom{M}{N} p^N (1-p)^{M-N}$ for $M$ channels with \textit{identical} transmission probabilities $p$, which, in turn, is composed of $M$ Bernoulli distributions $P_N^{\text{\,Bernoulli}}=(1-p)\,\delta_{N,0} + p\,\delta_{N,1}$ with success probability $p$, by allowing for pairwise \textit{different} transmission probabilities $p_i \neq p_j$ for $i\neq j$ and an infinite number of channels, $M \to \infty$.

In order to establish an experimentally feasible  procedure for detecting the deviation from a generalized binomial distribution, an unambigous criterion (analogous to the violation of the Bell inequalities to demonstrate entanglement) is needed.
This is, in the pioneering work of Kambly \textit{et al.} \cite{kamblyFactorial2011}, provided by proposing the use of so-called
\textit{factorial cumulants} 
    \begin{align}
    C_{\text{F},m}:=\left.\partial_z^m \mathcal{S}_\text{F}(z)\right|_{z=1}
\end{align}
as derivatives of the cumulant generating function
\begin{align}
    \mathcal{S}_\text{F}(z):=\ln \sum_{N}z^NP_N =: \ln \mathcal{M}_\text{F}(z)
    \label{eq:generating function}
\end{align}
with respect to the complex variable $z$.
For generalized binomial distributions, $\mathcal{S}_\text{F}(z)$ reduces to the sum
\begin{align}
    \mathcal{S}_\text{F}(z)=\sum_j\ln (1-p_j+z\,p_j)
    \label{eq:generating function generalized binomial}
\end{align}
over the Bernoulli contributions for channel $j$.
As a consequence, $\mathcal{S}_\text{F}(z)$ has singularities at the zeros $z_j=1-\frac{1}{p_j}$ of the moment generating function $\mathcal{M}_\text{F}(z)$. 
Because of $p_j \in [0,1]$, these zeros all lie on the negative real axis, $z_j  \in \mathbb{R}^{\le0}$.
Using $\ln(1+x)=\sum_{n=1}^\infty (-1)^{n-1}\frac{x^n}{n}$ with $x=(z-1)p_j$, it is straightforward to show \cite{kamblyFactorial2011,stegmannShorttime2016} that
\begin{align}
    C_{\text{F},m}=(-1)^{m-1}(m-1)!\sum_{j}p_j^m \, .
    \label{eq:factorial cumulants generalized binomial}
\end{align}
For a generalized binomial distribution with $p_j \in [0,1]$, the factorial cumulants obey the sign rule \cite{kamblyFactorial2011,stegmannDetection2015,stegmannShorttime2016}
\begin{align}
    (-1)^{m-1}C_{\text{F},m}\ge0
    \label{eq:sign rule}
\end{align}
as well as the normalization rule $\frac{(-1)^{m-1}}{(m-1)!} \frac{C_{\text{F},m}}{C_{\text{F},1}} \le 1$ \cite{gluthUnpublished}.

Not every distribution $P_N$ is of generalized binomial type.
Nevertheless, every $P_N$ with a finite support $0 \le N \le N_\text{max}$ can still be written in the form of Eq.~\eqref{eq:generalized binomial} with factorial cumulants given by Eq.~\eqref{eq:factorial cumulants generalized binomial}
by defining $p_j:=1/(1-z_j)$ through the, in general, complex-valued zeros $z_j\in \mathbb{C}$ of $\mathcal{M}_\text{F}(z)$, see also Appendix~\ref{sec: Representation of arbitrary discrete probability distributions}.
For $p_j \notin [0,1]$ (when $z_j$ lies \textit{outside} the negative real axis and $p_j$ can no longer be interpreted as a transmission probability), the sign (or the normalization) rule may be violated.

The Abanov-Ivanov theorem implies that for such a violation the presence of electron-electron interaction is necessary (though not sufficient).
Various systems have been proposed in which a sign-rule violation for fermionic counting statistics is expected to occur \cite{kamblyFactorial2011,stegmannDetection2015,kleinherbersRevealing2018,kleinherbersFull2023,zollnerDark2024}, but no experimental realization has been demonstrated so far.
In this work, we report on the first measurement where this is achieved in a tuneable manner.

What is the simplest possible model where a sign-rule violation of the factorial cumulants is expected?
As a starting point, we consider a single tunnel junction, where uncorrelated tunnel events occur with rate $\Gamma$.
The full counting statistics $P_N$ for a given time interval of length $t$ is obtained by dividing the time interval into $M \gg 1$ segments with a charge-transfer probability of $p=\Gamma t/M$ for each segment.
The resulting binomial distribution converges in the limit $M\to\infty$ into the Poisson distribution $P_N^{\text{\,Poisson}}= (\Gamma t)^N e^{-\Gamma t}/N!$ with the cumulant generating function $\mathcal{S}_\text{F}^{\text{\,Poisson}}(z)= (z-1) \Gamma t$, such that all higher-order factorial cumulants vanish, $C_{\text{F},m}=0$ for $m\ge 2$.

As a next step, we introduce an internal degree of freedom, namely a single \textit{spinless} orbital in the scattering region.
Then, charge transfer becomes correlated due to the Pauli principle: once an electron has entered the orbital, a second electron can only enter after the first electron has left.
These correlations yield non-vanishing factorial cumulants that, however, fulfill the sign rule, see Appendix~\ref{sec:Uncorrelated transport in two state systems}.
The minimal model allowing for a violation of the sign rule consists of a single \textit{spinful} orbital with an infinite charging energy that completely forbids double occupancy.
The orbital can, thus, be empty or singly charged with a spin-up or a spin-down electron, but not doubly charged. This three-state situation is a necessary, though not sufficient, condition. 
Whether or not the sign rule is actually violated depends on the system parameters.
In the present study, an external magnetic field is the decisive parameter that controls the appearance of a sign-rule violation.

To simplify the model further, the single orbital can be coupled to \textit{one} electron reservoir only, such that charging and discharging happens with the same lead.
The statistics for the discharging events is then equivalent to the statistics of unidirectional charge transfer in a two-terminal setup with properly-chosen parameters of the leads and the lead-orbital couplings, such that the general discussion outlined above applies.

\section{Experiment}

\begin{figure}
    \centering
    \includegraphics[width=0.9\linewidth]{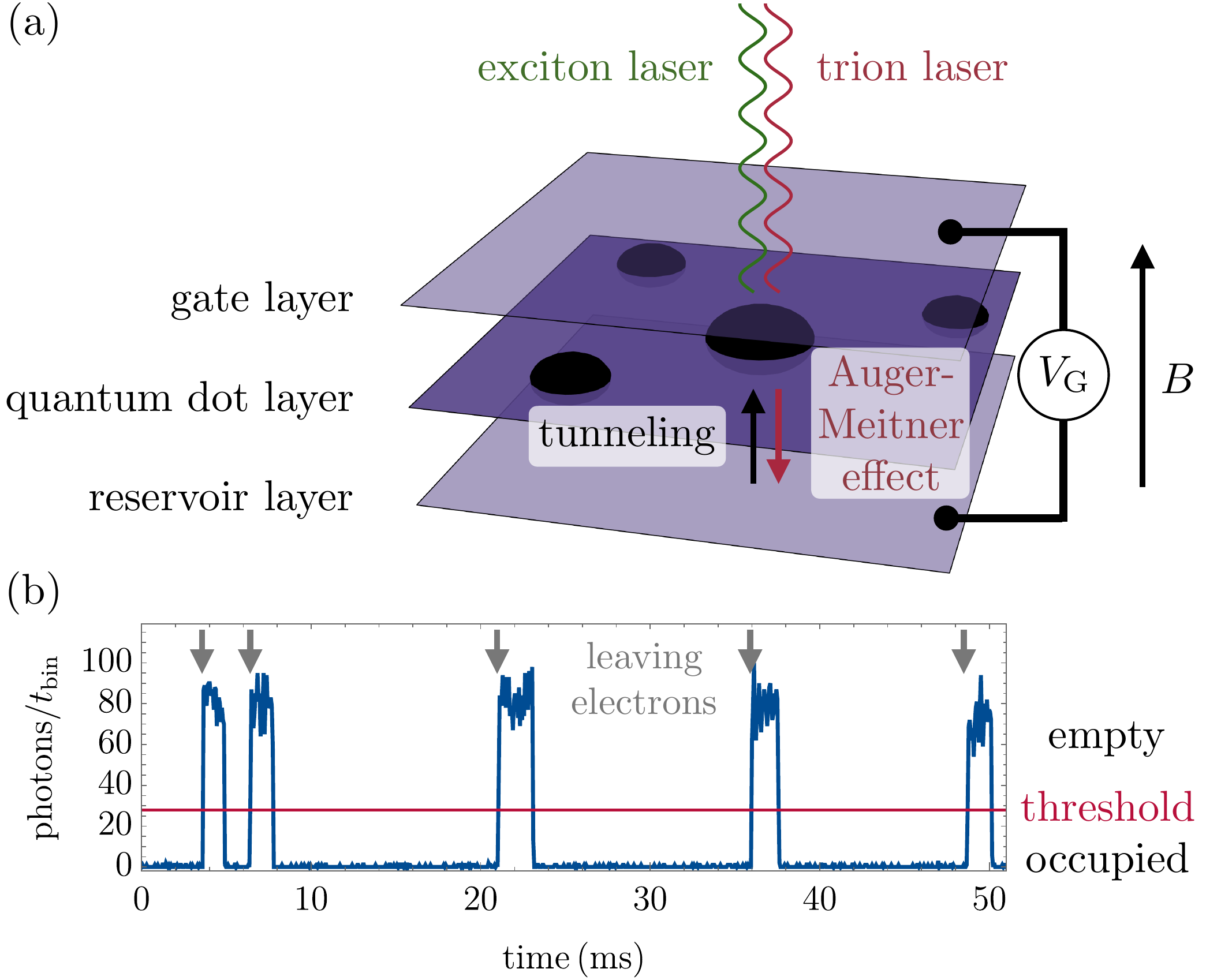}
    \caption{
    (a) The sample consists of a gate layer, the quantum-dot layer containing the investigated quantum dot, and a reservoir layer. The states of the quantum dot can be manipulated by gate voltage $V_\text{G}$ and magnetic field $B$. Electrons tunnel into the quantum dot (black arrow) and leave it via Auger processes caused by the trion laser (red). The occupation of the quantum dot is tracked using resonance fluorescence induced by the exciton laser (green).
    (b) Telegraph signal resulting from detected photons, which indicates the occupation of the quantum dot. The number of photons detected within the interval $t_\text{bin}=60\,\mu$s is tracked over time.
    }
    \label{fig:setup}
\end{figure}

We realize this minimal model with an InAs quantum dot that is tunnel-coupled to a three-dimensional electron gas in a $p$-$i$-$n$ diode structure, as depicted in Fig.~\ref{fig:setup}(a). For details, see Refs.~\cite{mannelAuger2023,kleinherbersUnraveling2023} and references therein.
A voltage $V_\text{G}$ is applied such that the quantum dot is either empty or singly occupied with a spin-up or spin-down electron.
The small size of the dot implies a large charging energy, which prohibits double occupancy.  
The sample is subject to a tuneable magnetic field $B$, which lifts the spin degeneracy.
The occupation of the quantum dot is monitored in real time via resonance fluorescence, as demonstrated in Ref.~\cite{kurzmannOptical2019}:  
Laser light resonantly drives transitions between the ground state of the uncharged quantum dot and an exciton state, i.e. a bound electron-hole pair. 
This leads to a resonance fluorescence signal of the exciton transition.
When an electron tunnels into the quantum dot, the resonance condition is no longer met, and the fluorescence turns off.
In the bright state, the emitted photons, detected by an avalanche photo diode, are recorded and a posteriori binned to obtain the telegraph signal \cite{kurzmannOptical2019,kerskiPostprocessing2023,kleinherbersPushing2022} shown in Fig.~\ref{fig:setup}(b). The onsets of the fluorescence (gray arrows) are used to identify the times when an electron is ejected out of the dot. 

As an alternative to the resonance-fluorescence-based method used here \cite{efrosRandom1997,vamivakasObservation2010,kurzmannOptical2019,kleinherbersUnraveling2023,kerskiPostprocessing2023,sifftRevealing2025,mannelQuantum2025}, the occupation of a quantum dot can also be detected electrically via a capacitative coupling to another quantum dot or a quantum point contact \cite{gustavssonCounting2006,fujisawaBidirectional2006,flindtUniversal2009,frickeHigh2010,komijaniCounting2013,heckerRadioFrequency2025,garreisCounting2023}.
Counting quantum jumps from telegraph signals has proven to be a powerful tool to gain an understanding of tunneling dynamics in quantum dots \cite{gustavssonCounting2006, fujisawaBidirectional2006,komijaniCounting2013,rudgeCounting2019,kleinherbersUnraveling2023,mannelQuantum2025} and photon statistics \cite{mandelFluctuations1959,mandelSubPoissonian1979,beenakkerCounting2001,brangePhoton2019,beenakkerThermal1998,zenelajFull2022}.

In the present system, the magnetic field turns out to be a control parameter for the violation of the sign rule.
At $B=0$, spin degeneracy implies that 
the charge dynamics can be effectively described within a two-state model for a charged and an uncharged quantum dot.
Despite of the electron-electron interaction, the sign rule is, nevertheless, fulfilled since this is the case for \textit{any} two-state model, as shown in Appendix~\ref{sec:Uncorrelated transport in two state systems}.
Consequently, at least \textit{three} states are needed for a sign-rule violation.
This can be achieved by lifting the spin degeneracy with a magnetic field.
With increasing magnetic field the spin-relaxation rate increases.
When the population of the excited spin state is strongly suppressed by spin relaxation, the system can again be described by an effective two-state model.
But even when the three-state model applies, a sign-rule violation is not guaranteed \cite{kleinherbersUnraveling2023}.

\begin{figure*}
    \centering
    \includegraphics[width=.65\linewidth]{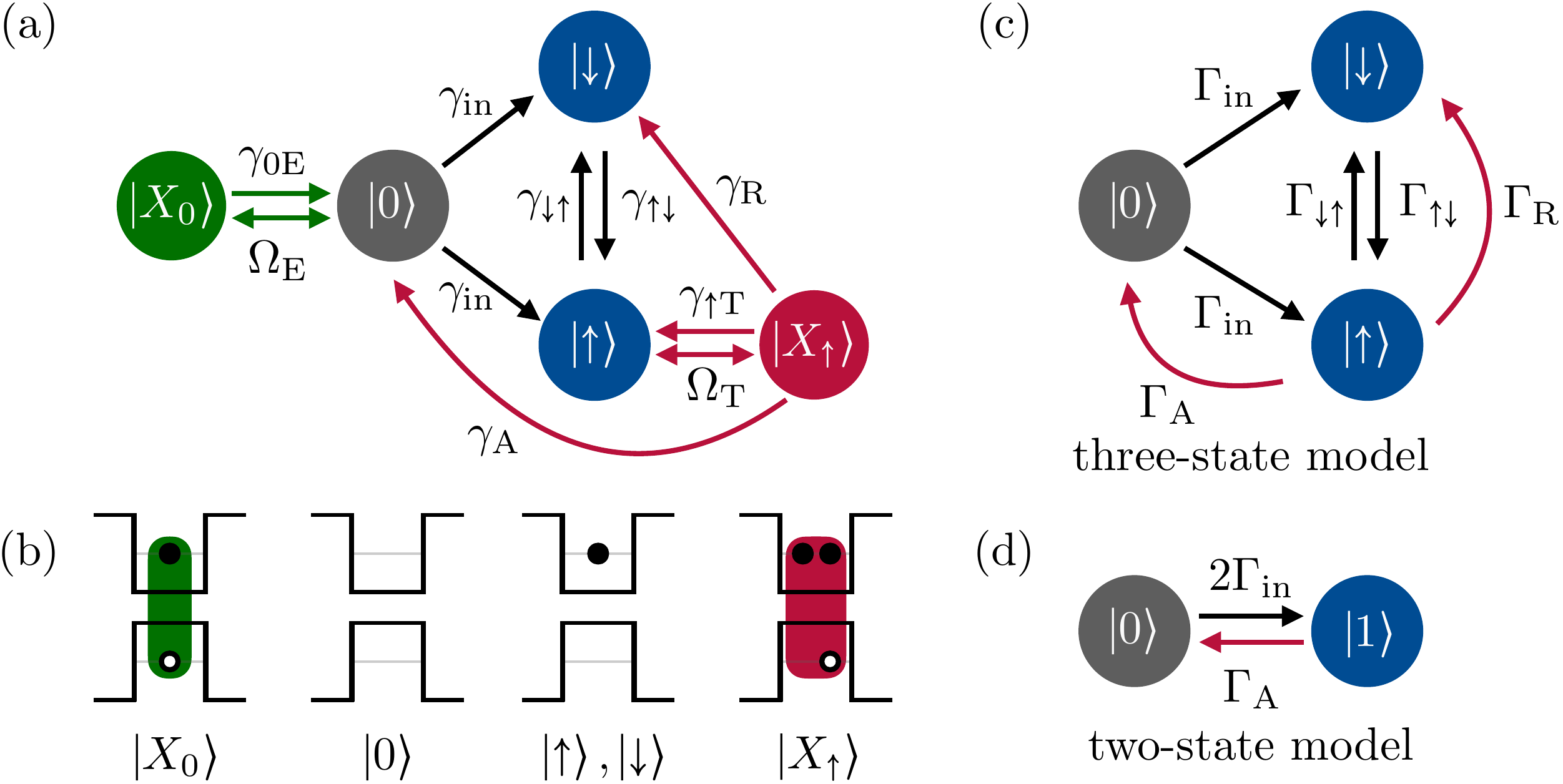}
    \caption{(a) Relevant states and possible transitions between them. (b) Particles involved for each state with electrons represented by filled circles and holes represented by empty circles. The confining potential is sketched as a rectangular quantum well for the electron and as an inverted quantum well for the hole.  (c) Simplified three-state model. Exciton state $\ket{X_0}$ and unoccupied state $\ket{0}$ as well as trion state $\ket{X_\uparrow}$ and spin-up state $\ket{\uparrow}$ have been combined. 
    (d) Two-state model, applicable for very small or very large magnetic fields.}
    \label{fig:transitions}
\end{figure*}

In Fig.~\ref{fig:transitions}(a), we show a model that has been used successfully to describe all relevant transitions in our system \cite{mannelAuger2023,kleinherbersUnraveling2023}.
As discussed above, one laser ("exciton laser") resonantly drives transitions between the unoccupied quantum dot $\ket{0}$ and the exciton $\ket{X_0}$. The emitted photons are used to monitor the charge state of the quantum dot. 
Electrons can tunnel from the reservoir layer into the spin states $\ket{\downarrow}$ and $\ket{\uparrow}$.
The corresponding rate is $\gamma_\text{in}$ for each spin.
Within the charged region, spin relaxation \cite{kroutvarOptically2004,atatureQuantumDot2006,luDirect2010,mannelAuger2023,kleinherbersUnraveling2023} from state $\ket{\downarrow}$ to $\ket{\uparrow}$ occurs with rate $\gamma_{\uparrow\downarrow}$. 
The reverse process is less likely.
Its rate $\gamma_{\downarrow\uparrow}=\gamma_{\uparrow\downarrow}\exp \left(-\frac{\Delta_e}{k_\text{B}T} \right)$ depends on the temperature and on the magnetic field $B$ via the Zeeman energy $\Delta_e(B)$. 

For the chosen gate voltage, tunneling out of the dot is energetically forbidden.
Instead, we use the Auger-Meitner effect \cite{kurzmannAuger2016} to discharge the quantum dot. 
A second laser ("trion laser") resonantly drives transitions between $\ket{\uparrow}$ and the trion state $\ket{X_{\uparrow}}$, which is formed by the electron of the charged quantum dot and an additional electron-hole pair, with rate $\Omega_\text{T}$ (see Figs.~\ref{fig:transitions}(a) and (b)). 
The driving of the trion resonance leads to additional processes.
On one hand, it enables the Auger-Meitner effect, in which the energy from the electron-hole recombination is used to eject the remaining electron from the quantum dot with rate $\gamma_\text{A}$ \cite{efrosRandom1997,kurzmannAuger2016,loblRadiative2020,gillardFundamental2021,mannelAuger2023,kleinherbersUnraveling2023}. 
On the other hand, the trion state $\ket{X_\uparrow}$ can decay into the excited spin state $\ket{\downarrow}$ with rate $\gamma_\text{R}$, a process referred to as spin-flip Raman scattering \cite{atatureQuantumDot2006,debusSpinflip2014,mannelAuger2023,kleinherbersUnraveling2023}. 
In addition, there is a spontaneous decay from state $\ket{X_\uparrow}$ to the ground state $\ket{\uparrow}$ with rate $\gamma_{\uparrow\text{T}}$. 
Both the rate of the Auger-Meitner process $\gamma_\text{A}$ and the rate of the spin-flip Raman scattering $\gamma_\text{R}$ depend on the trion laser intensity. For all measurements presented here, the trion laser intensity is tuned in such a way, that it enables a sufficient statistics of entering and leaving electrons, see Table~\ref{tbl:transition rates}.
The processes induced by the trion laser are represented by red arrows in Fig.~\ref{fig:transitions}(a).

\section{Model}

The full model contains the five states $\ket{0}$, $\ket{\uparrow}$, $\ket{\downarrow}$, $\ket{X_0}$, and $\ket{X_\uparrow}$.
Since the laser-driven (exciton and trion) transitions $\ket{0}\leftrightarrow \ket{X_0}$ and $\ket{\uparrow} \leftrightarrow \ket{X_\uparrow}$ happen on a much faster time scale than the other (charge and spin) transitions, the states connected by these  optically-driven transitions can be combined into new effective states. 
This results in an effective three-state model \cite{kleinherbersUnraveling2023}, as shown in Fig.~\ref{fig:transitions}(c), with effective rates denoted by capital ($\Gamma$) instead of lowercase ($\gamma$) letters.
In the limits of either a very small or a very large magnetic field, the system simplifies to an effective two-state model (see Fig.~\ref{fig:transitions}(d)), in the first case because the two spin states $\ket{\uparrow}$ and $\ket{\downarrow}$ are equally populated, in the second case because spin relaxation is instantaneous so that only the ground-state $\ket{\uparrow}$ has to be taken into account.

To calculate the cumulant generating function $\mathcal{S}_\text{F}(z,t)$ as a function of the length $t$ of the time interval, in which discharging events are counted, we consider the state- and $N$-resolved master equation 
\begin{align}
    \dot{\boldsymbol{P}}_N(t)=\left(\boldsymbol{W}-\boldsymbol{\mathcal{J}}_\text{out}\right)\boldsymbol{P}_N(t)+\boldsymbol{\mathcal{J}}_\text{out}\boldsymbol{P}_{N-1}(t) \, ,
    \label{eq:differential equation}
\end{align}
written in matrix-vector notation.
Here, ${\boldsymbol{P}_N=(P_N^{0},P_N^\downarrow,P_N^\uparrow)}^T$ is the vector of the probabilities $P_N^\chi$ that the quantum dot is in state $\chi$ and $N$ discharging events have been counted.
The off-diagonal elements of the matrix
\begin{align}
    \boldsymbol{W}=
    \begin{pmatrix}
        -2\Gamma_\text{in} && 0 &&\Gamma_\text{A}\\
        \Gamma_\text{in} && -\Gamma_{\uparrow\downarrow} &&\Gamma_\text{R}+\Gamma_{\downarrow\uparrow}\\
        \Gamma_\text{in} && \Gamma_{\uparrow\downarrow} && -\Gamma_\text{A}-\Gamma_\text{R}-\Gamma_{\downarrow\uparrow}
    \end{pmatrix}
    \label{eq:W3}
\end{align}
describe the transition rates, and the diagonal matrix elements are determined from the condition $\mathbf{e}^T \mathbf{W}=\mathbf{0}$ with $\mathbf{e}^T:=(1,1,1)$ and $\mathbf{0}^T:=(0,0,0)$, which guarantees the normalization $\sum_\chi \sum_N P_N^\chi=1$.
The jump operator
\begin{align}
    \boldsymbol{\mathcal{J}}_\text{out}=
    \begin{pmatrix}
        0&&0&&\Gamma_\text{A}\\
        0&&0&&0\\
        0&&0&&0
    \end{pmatrix}
    \label{eq:J3_out}
\end{align}
singles out the transition that is counted (here discharging the dot due to the Auger-Meitner effect).
After performing the $z$-transform $\mathbf{P}_z := \sum_N z^N \mathbf{P}_N$, the master equation becomes
$\dot{\boldsymbol{P}}_z(t)=\left[ \boldsymbol{W}+(z-1)\boldsymbol{\mathcal{J}}_\text{out} \right] \boldsymbol{P}_z(t)$, which is formally solved by $\boldsymbol{P}_z(t)=\exp(\mathbf{W}_z t) \boldsymbol{P}_\text{st}$ with $\boldsymbol{W}_z:= \boldsymbol{W}+(z-1)\boldsymbol{\mathcal{J}}_\text{out}$ and the stationary probability vector $\boldsymbol{P}_\text{st}$ obtained from $\boldsymbol{W}\boldsymbol{P}_\text{st}=0$ and $\boldsymbol{e}^T\boldsymbol{P}_\text{st}=1$.
Multiplying the solution of the master equation from the left with $\boldsymbol{e}^T$ yields 
\begin{align}
\mathcal{S}_{\text{F}}(z,t)= \ln \left[ \boldsymbol{e}^T \exp(\mathbf{W}_z t) \boldsymbol{P}_{\text{st}}\right] \, ,
\label{eq:generating function solution}
\end{align}
which we use for the numerical evaluation of the factorial cumulants.
For the two-state model, we have to use $\mathbf{e}^T:=(1,1)$ and $\mathbf{0}^T:=(0,0)$, as well as
\begin{align}
    \boldsymbol{W}=
    \begin{pmatrix}
        -2\Gamma_\text{in} &&\Gamma_\text{A}\\
        2\Gamma_\text{in} && -\Gamma_\text{A}
    \end{pmatrix} 
    \;\; \text{and} \;\;
    \boldsymbol{\mathcal{J}}_\text{out}=
    \begin{pmatrix}
        0&&\Gamma_\text{A}\\
        0&&0
    \end{pmatrix}.
        \label{eq:W2}
\end{align}
To model a limited time resolution of the detector, which is given by the binning time $t_\text{bin}$ of the photon counts, we discretize the master equation following Ref.~\cite{kleinherbersPushing2022}, see Appendix~\ref{sec:Influence of the limited time resolution on the factorial cumulants}.

\section{Results and discussion}

\begin{figure*}
    \centering
    \includegraphics[width=\linewidth]{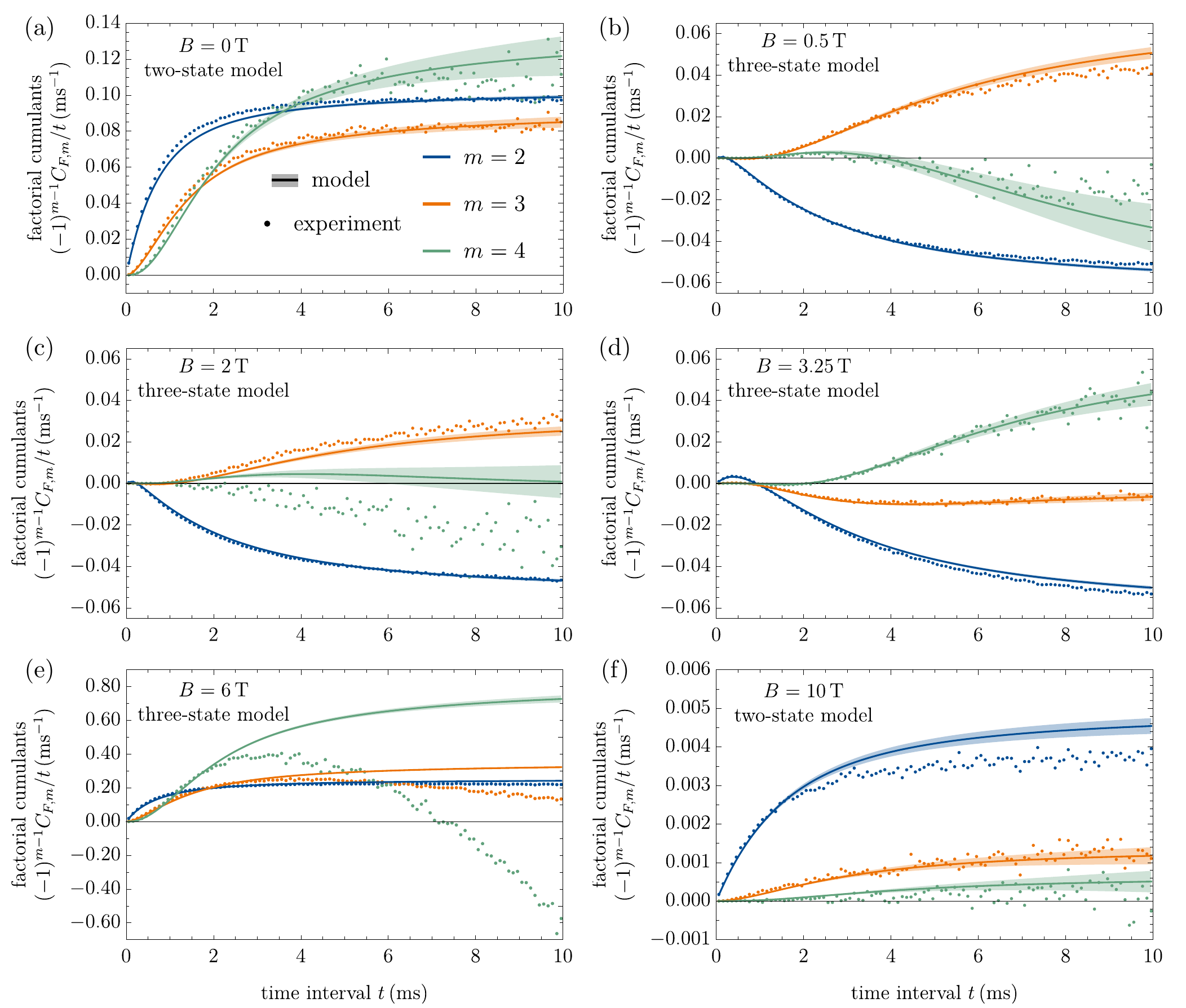}
    \caption{Factorial cumulants as a function of the time interval for magnetic fields (a) $B=0\,$T, (b) $B=0.5\,$T, (c) $B=2\,$T, (d) $B=3.25\,$T, (e) $B=6\,$T, and (f) $B=10\,$T. The experimental data is shown as dots, the model calculations as straight lines and the expected standard deviation from the model as colored regions. 
    In (b)-(d), some of the factorial cumulants change sign, while following the sign rule \eqref{eq:sign rule} in (a) and (f). 
    }
    \label{fig:cumulants}
\end{figure*}

We measure the charge dynamics in our quantum dot at temperature $T=4.2\,$K and six different values of the magnetic field.
For this, the stream of detected photons is binned using a binning time of $t_\text{bin}=60\,\mu$s to generate random telegraph signals like the one shown in Fig.~\ref{fig:setup}(b).
Each stream has a length of about $900\,$s, comprising roughly $10^5$ -- $10^6$ events. 
From the random telegraph stream, we extract the second-, third-, and fourth-order factorial cumulants. They are displayed in Fig.~\ref{fig:cumulants} as blue, orange, and green data points, respectively for (a) $B=0\,$T, (b) $B=0.5\,$T, (c) $B=2\,$T, (d) $B=3.25\,$T, (e) $B=6\,$T, and (f) $B=10\,$T as a function of the time interval $t$, during which discharging events are counted.
The factorial cumulants $C_{\text{F},m}(t)$ are divided by $t$, such that each data set approaches a constant in the long-time limit, and multiplied with the factor $(-1)^{m-1}$, such that the sign-rule violation is indicated by \textit{negative} values.

As displayed in Fig.~\ref{fig:cumulants}(b)-(d), there is a pronounced violation of the sign rule for some of the higher-order factorial cumulants, which is the main experimental result of our work.
In contrast, at $B=0\,$T and $B=10\,$T the sign rule is fulfilled despite the presence of electron-electron interaction. Here, the system can still be effectively described in terms of single-particle charge transfers.

To validate our interpretation, we model the system within the rate-equation approach described above.  
For this, we extract the effective rates as depicted in Fig.~\ref{fig:transitions}(c) and (d), from the random telegraph signal in the same way as in Ref.~\cite{kleinherbersUnraveling2023}, see Appendix~\ref{sec:Extracting tunneling rates from waiting time distribution}. 
For $B=0\,$T and $B=10\,$T, we use a two-state model that is consistent with the monoexponential behavior of the measured waiting-time distribution, while for the other magnetic fields, a three-state model must be used to explain the biexponential decay, see Fig.~\ref{fig:waitingtimes}.
We also include the limited time resolution of the detection scheme, that is due to the binning of the photon counts, see Appendix~\ref{sec:Influence of the limited time resolution on the factorial cumulants}.
The theoretical results are shown in Fig.~\ref{fig:cumulants} as solid lines.
To estimate statistical errors due to the finite length of the random telegraph stream,  we calculated the expected standard deviation for each factorial cumulant, shown as the colored areas, following Refs.~\cite{kleinherbersPushing2022,kleinherbersUnraveling2023}.

For all magnetic fields, except for $B=6\,$T, experimental data and theory in Fig.~\ref{fig:cumulants} show a good agreement, which strongly supports our interpretation and gives us confidence that our model correctly describes the experimental situation. In particular, it justifies the reduction of the complete model (Fig.~\ref{fig:transitions}(a)) to the simple three-state version (Fig.~\ref{fig:transitions}(c)), which constitutes the simplest situation, where mapping of the interacting system onto a single-particle description may no longer be possible.
The slight deviation between theory and experiment for the second factorial cumulant at $B=10\,$T, see Fig.~\ref{fig:cumulants}(f), indicates that the assumption of instantaneous spin relaxation may not be fully justified.

At $B=6\,$T, the biexponential waiting-time distribution shown in Fig.~\ref{fig:waitingtimes}(e) indicates a three-state model.
Nevertheless, the calculated factorial cumulants fulfill the sign rule.
We consider the deviation between experimental data and modeling, including the observed sign change of the fourth factorial cumulant, as an artifact of the measurement.
A direct inspection of the telegraph signal indicates a bistability (of unknown origin) of the system with a switching time of the order of 100 ms. Another clear example for this can be found in Fig.~5 and 6 of Ref.~\cite{kleinherbersUnraveling2023} which show a biexponential waiting time distribution, i.e., three-state behavior, but no violation of the sign rule.

We conclude the discussion with two important remarks.
First, we emphasize that having a three- instead of a two-state model is necessary but not sufficient for a sign-rule violation of the factorial cumulants.
At moderately high fields, we find a biexponential waiting-time distribution but the corresponding calculated factorial cumulants fulfill the sign rule.

Second, the $(-1)^{m-1}$ term in the sign rule Eq.~(\ref{eq:sign rule}) is relevant for \textit{fermionic} counting statistics only, not for \textit{bosons}.
In fact, the factorial cumulants for photons emitted in a microwave cavity are predicted to be positive \cite{brangePhoton2019} for all orders.
The measured charge-transfer statistics of Cooper pairs in a metallic single-electron box has been understood in terms of a Poisson distribution of avalanches of Andreev processes, which would yield positive factorial cumulants \cite{maisiFull2014}. 
Furthermore, the factorial cumulants for coherent spin excitations in carbon-nanotube quantum dots functionalized with a sizable number of molecular antiferromagnets have been found to be positive \cite{bessonIndependent2023}.
In all these cases, the violation of the \textit{fermionic} sign rule is simply due to the \textit{bosonic} nature of the counted objects.

\section{Conclusions}

In conclusion, we experimentally demonstrated the violation of the sign rule for the factorial cumulants of electron tunneling in a quantum dot.
This shows that the counting statistics cannot be of generalized binomial type and, therefore, not be understood within an effective single-particle picture.
This unambiguously proves the presence of correlations induced by electron-electron interaction.

We have investigated the simplest conceivable model system in which a violation of the fermionic sign rule is possible -- an arrangement of three different states with transitions between them.
The experimental realization consists of a single spinful orbital in a quantum dot, where an external magnetic field lifts the spin degeneracy. Both experiment and the theoretical model clearly show a violation of the sign rule for an intermediate range of magnetic fields.

With the present work we aim at establishing the measurement of factorial cumulants as a useful and feasible experimental tool to unambiguously prove interaction-induced correlations in the full counting statistics of fermionic charge transfer.
We advocate that it should become a standard experimental tool, since the violation of the sign rule provides new information that cannot be deduced from conventional statistical quantities, such as the waiting-time distribution. 

For future studies, alternative realizations in different material systems, setups, and measurement schemes for the charge detection are desirable.
In general, to find a sign-rule violation of the factorial cumulants, one should measure the statistics of electron transfer through a scattering region in the presence of electron-electron interaction. 
In a two-state model, the sign rule is always fulfilled.
Therefore, the minimal model should involve at least three states.
Since only some of all possible three-state models will show a sign-rule violation, the possibility to tune one or several system parameters would be advantageous.

\begin{acknowledgments}
This work was supported by the Deutsche Forschungsgemeinschaft (DFG, German  Research Foundation) under Project-ID 278162697 -- SFB 1242 and the Mercator Research Center Ruhr (MERCUR) under Project No. Ko-2022-0013. MG acknowledges the
individual research grant No. GE2141/5-1.
\end{acknowledgments}

\section*{Data Availability}
The data that support the findings of this article are openly available \cite{zollnerData}.

\appendix

\section{Extracting tunneling rates from waiting-time distributions}
\label{sec:Extracting tunneling rates from waiting time distribution}

The distribution $w(\tau)$ of times $\tau$ that pass between two types of events are called waiting-time distributions. They allow to characterize the quantum mechanical switching between states, which is tracked by telegraph signals, as shown in Fig.~\ref{fig:setup}(b) \cite{brandesWaiting2008,rudgeDistribution2018,brangeControlled2021,kleinherbersUnraveling2023,fussangelStatistical2025,sifftRevealing2025}.
We consider the waiting-time distribution $w_\text{occ}(\tau)$, which describes how long the quantum dot is occupied, and $w_\text{emp}(\tau)$, which describes how long the quantum dot is unoccupied.
They can be calculated using
\begin{align}
    w_\text{emp}(\tau) &= \frac{\boldsymbol{e}^T\boldsymbol{\mathcal{J}}_\text{in} e^{(\boldsymbol{W}-\boldsymbol{\mathcal{J}}_\text{in}-\boldsymbol{\mathcal{J}}_\text{out})\tau}\boldsymbol{\mathcal{J}}_\text{out}\boldsymbol{P}_\text{st}}{\boldsymbol{e}^T\boldsymbol{\mathcal{J}}_\text{out}\boldsymbol{P}_\text{st}} \, ,
    \label{eq:wemp}
    \\
    w_\text{occ}(\tau) &= \frac{\boldsymbol{e}^T\boldsymbol{\mathcal{J}}_\text{out} e^{(\boldsymbol{W}-\boldsymbol{\mathcal{J}}_\text{in}-\boldsymbol{\mathcal{J}}_\text{out})\tau}\boldsymbol{\mathcal{J}}_\text{in}\boldsymbol{P}_\text{st}}{\boldsymbol{e}^T\boldsymbol{\mathcal{J}}_\text{in}\boldsymbol{P}_\text{st}} \, .
    \label{eq:wocc}
\end{align}
where the matrix $\boldsymbol{W}$ is given by either Eq.~(\ref{eq:W3}) for the three-state or Eq.~(\ref{eq:W2}) for the two-state model.
The jump operator $\boldsymbol{\mathcal{J}}_\text{out}$ keeps from $\boldsymbol{W}$ only the off-diagonal matrix element containing $\Gamma_\text{A}$, see Eqs.~(\ref{eq:J3_out}) and (\ref{eq:W2}).
Similarly, $\boldsymbol{\mathcal{J}}_\text{in}$ keeps from $\boldsymbol{W}$ only the off-diagonal matrix elements containing $\Gamma_\text{in}$.

\begin{figure*}
    \centering
    \includegraphics[width=0.85\linewidth]{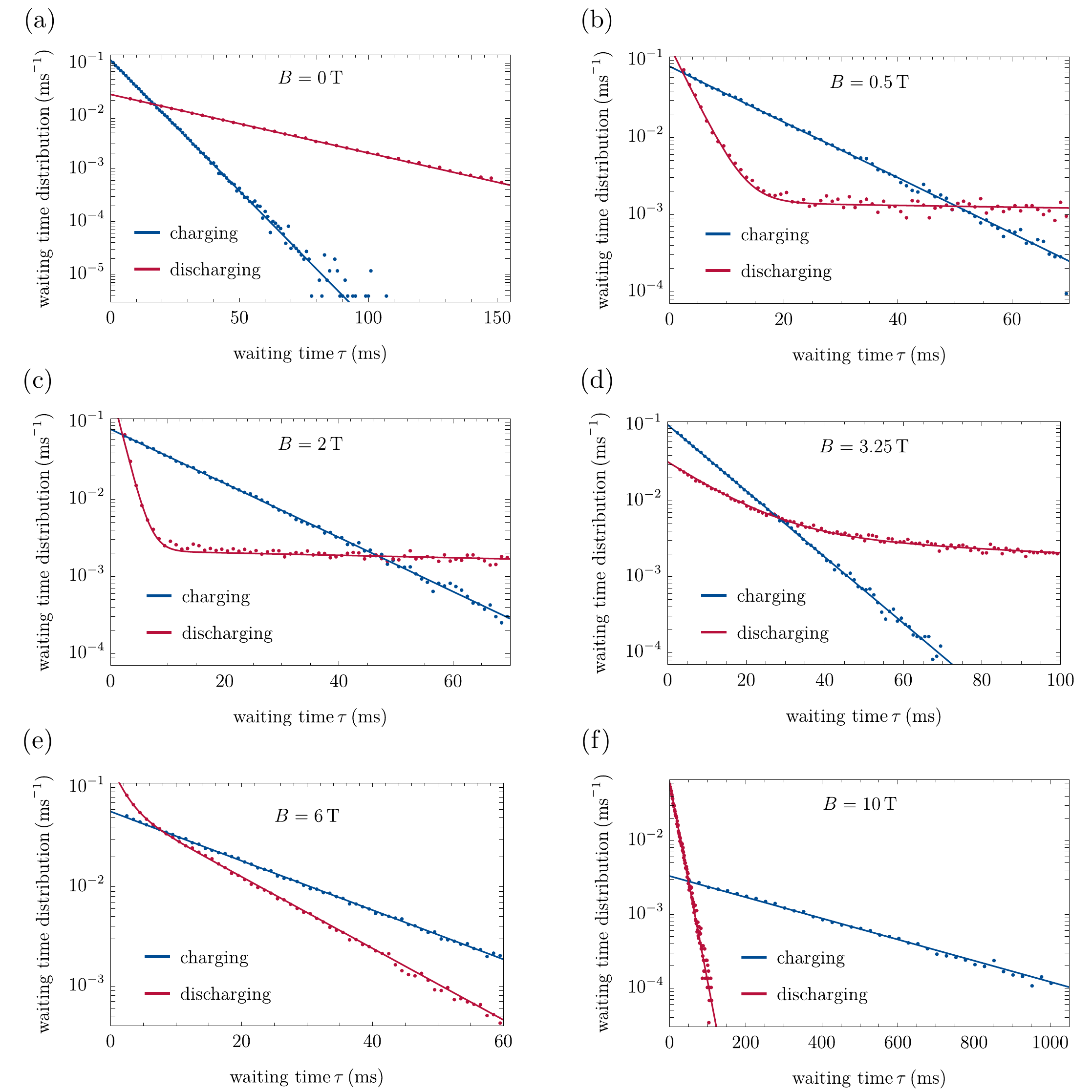}
    \caption{Waiting-time distributions $w_\text{emp}(\tau)$ (blue) and $w_\text{occ} (\tau)$ (red) at magnetic fields (a) $B=0\,$T, (b) $B=0.5\,$T, (c) $B=2\,$T, (d) $B=3.25\,$T, (e) $B=6\,$T, and (f) $B=10\,$T. Fitting the experimental waiting-time distributions (dots) with the expressions from Eq.~\eqref{eq:wemp 3s} and Eq.~\eqref{eq:wocc 3s} (lines) yields the transition rates presented in Table~\ref{tbl:transition rates}.
    }
    \label{fig:waitingtimes}
\end{figure*}

The waiting-time distributions for the full three-state model resulting from Eq.~\eqref{eq:wemp} and Eq.~\eqref{eq:wocc} are given by
\begin{align}
    w_\text{emp}(\tau) &= 2\Gamma_\text{in}e^{-2\Gamma_\text{in}\tau},
    \label{eq:wemp 3s}
    \\
    w_\text{occ}(\tau) &= \alpha\Gamma_1 e^{-\Gamma_1 \tau}+(1-\alpha)\Gamma_2 e^{-\Gamma_2 \tau}.
    \label{eq:wocc 3s}
\end{align}
While $w_\text{emp}(\tau)$ is always monoexponential, $w_\text{occ}(\tau)$ is more complex. 
The dependence of the parameters $\alpha$, $\Gamma_1$ and $\Gamma_2$ of the transition rates $\Gamma_\text{A}$, $\Gamma_{\uparrow\downarrow}$ and $\Gamma_\text{R}$ is complicated and given by Eq.~(11) of Ref.~\cite{kleinherbersUnraveling2023}. 
The remaining transition rate $\Gamma_{\downarrow\uparrow}=\Gamma_{\uparrow\downarrow}\exp\left(-\frac{\Delta_e}{k_\text{B}T}\right)$ can be calculated from the temperature $T=4.2\,$K and the Zeeman energy $\Delta_{e}$ which is a function of the magnetic field $B$ (see Table~\ref{tbl:transition rates}).

For very large or very small magnetic fields, the model reduces to a two-state model.
The distribution for $w_\text{occ}(\tau)$ is, then, monoexponential with rate $\Gamma_\text{A}$, which can be formally obtained from Eq.~\eqref{eq:wocc 3s} by setting $\alpha=1$ and $\Gamma_1=\Gamma_\text{A}$.
The biexponential distribution Eq.~\eqref{eq:wocc 3s} only occurs for intermediate magnetic fields.

In Fig.~\ref{fig:waitingtimes}, we show the waiting-time distribution extracted from the telegraph signals for the six magnetic fields.
By fitting Eqs.~\eqref{eq:wemp 3s} and \eqref{eq:wocc 3s}, we obtain the transition rates listed in Table~\ref{tbl:transition rates}.
The waiting times for magnetic fields from $B=0.5\,$T to $B=6\,$T where fitted using the full three state model.
It turns out that it becomes harder to separate the two components of the biexponential distribution $w_\text{occ}(\tau)$ with increasing magnetic field.
While the use of the two-state model is fully justified for $B=0\,$T by spin symmetry, there might be still a small but finite occupation of the state $\ket{\downarrow}$ for $B=10\,$T.
It is, however, futile to use the three-state model in this case since the fast spin-relaxation processes are not resolved by the waiting times.

\begin{table*}[htbp]
    \begin{tabular}{|c|c|c|c|c|c|c|c|}
        \hline
        $B\,$(T) &$\Gamma_\text{in}\,$(kHz)&$\Gamma_\text{A}\,$(kHz)&$\Gamma_{\uparrow\downarrow}\,$(kHz)&$\Gamma_{\downarrow\uparrow}\,$(kHz)&$\Gamma_\text{R}\,$(kHz) & $\Delta(k_\text{B}T)$& $I_\text{T}(I_0)$\\
        \hline
        $0$&$0.95$&$0.43$&-&-&-&-&$0.1$ \\
        \hline
        $0.5$&$0.69$&$5.14$&$0.049$&$0.046$&$0.63$ &$0.064$&$0.5$\\
        \hline
        $2$&$0.67$&$9.38$&$0.074$&$0.057$&$1.96$ &$0.256$&$0.5$ \\
        \hline
        $3.25$&$0.84$&$1.09$&$0.115$&$0.076$&$0.176$& $0.416$ & $1$\\
        \hline
        $6$&$0.48$&$5.69$&$2.09$&$0.97$&$1.22$& $0.768$& $1$\\
        \hline
        $10$&$0.52$&$0.055$&-&-&-&-&$1$ \\
        \hline
    \end{tabular}
    \caption{Transition rates, Zeeman energies and laser intensities for different magnetic fields.}
    \label{tbl:transition rates}
\end{table*}

\section{Two-state model}
\label{sec:Uncorrelated transport in two state systems}

In this appendix, we show that for any two-state model with transition rates $\Gamma_1$ and $\Gamma_2$ between the two states, the sign rule is obeyed for all factorial cumulants if only one of the two transitions is counted.
By introducing the average transition rate $\Gamma:=(\Gamma_1+\Gamma_2)/2$ and the asymmetry $a:=(\Gamma_1-\Gamma_2)/(\Gamma_1+\Gamma_2)$ (with $-1 \le a \le 1)$, the matrix governing the full counting statistics can be written in the form
\begin{align}
    \boldsymbol{W}_z=\Gamma\begin{pmatrix}
        -1-a & z(1-a)\\
        1+a & -1+a
    \end{pmatrix} \, .
\end{align}
Calculating $\boldsymbol{P}_\text{st}$
from $\boldsymbol{W}_1\boldsymbol{P}_\text{st}=0$ and $\boldsymbol{e}^T\boldsymbol{P}_\text{st}=1$, plugging $\boldsymbol{W}_z$ and $\boldsymbol{P}_\text{st}$ into Eq.~\eqref{eq:generating function solution}, and using Eq.~\eqref{eq:generating function} yields the moment generating function $\mathcal{M}_\text{F}(z,t)$.
It is straightforward to write it in the form (see also Ref.~\cite{kamblyFactorial2011})
\begin{align}
    \mathcal{M}_\text{F}(z,t)=\frac{e^{-\Gamma t}}{4\eta}
    \Biggl[\underbrace{\left(1+\eta\right)^2 e^{\Gamma t \eta}}_{=:M_+(z)}
    -\underbrace{\left(1-\eta\right)^2 e^{-\Gamma t \eta}}_{=:M_-(z)}\Biggr] 
\label{eq:moment_generating_function_2_state}
\end{align}
where $\eta(z)$ is one of the complex solutions of
\begin{align}
    \eta^2 = z(1-a^2)+a^2 \, .
    \label{eq:eta}
\end{align}
Since $\eta \to -\eta$ leaves $\mathcal{M}_\text{F}(z,t)$ invariant, it is irrelevant which of the two complex solutions is chosen.

To prove that the sign rule for the factorial cumulants is fulfilled, we need to show that, for all times $t$, all zeros $z_j$ of the moment generating function $\mathcal{M}_\text{F}(z,t)$ lie on the negative real axis.
From \eqref{eq:moment_generating_function_2_state}, we observe that the zeros $z_j$ are given by $M_+(z_j)=M_-(z_j)$, i.e., both the modulus and the phase of $M_+$ and $M_-$ need to match.
This leads to a transcendental equation for the zeros, which can be solved numerically \cite{kamblyFactorial2011}.
For our purpose, however, it is sufficient to analyze the necessary (but not sufficient) condition
\begin{align}
    \left| M_+(z_j) \right| = \left| M_-(z_j) \right|
\end{align}
for the zeros.
By splitting $\eta$ into its real and imaginary part, we rewrite the modulus of $M_\pm$ as
\begin{align}
    \left| M_\pm \right| =\left[\left(1\pm\text{Re}\, \eta \right)^2+\left(\text{Im}\, \eta\right)^2\right]e^{\pm \Gamma t\,  \text{Re}\, \eta} \, ,
\end{align}
and compare $\left| M_+ \right|$ with $\left| M_- \right|$.
For $\text{Re}\, \eta > 0$, both the exponential factor and the factor in brackets are larger for $\left| M_+ \right|$ than for $\left| M_- \right|$, which implies $\left| M_+ \right| >\left| M_- \right|$.
Similarly, for $\text{Re}\, \eta < 0$, we have $\left| M_+ \right| <\left| M_- \right|$.
As a consequence, $\left| M_+ \right| =\left| M_- \right|$ is fulfilled if and only if $\text{Re}\, \eta = 0$, which is the case if and only if the right hand side of Eq.~\eqref{eq:eta} lies on the negative real axis.

In conclusion, all zeros $z_j$ of $\mathcal{M}_\text{F}(z,t)$ lie within the real interval $( -\infty, -\frac{a^2}{1-a^2} ]$, which is, because of $0 \le a^2 \le 1$, a subset of the negative real axis.
The exact locations of the zeros within this interval depend on $a$ and $\Gamma t$.
They are determined by the additional condition that not only the modulus but also the phase of $M_+$ and $M_-$ have to match.
But in any case, all zeros $z_j$ of $\mathcal{M}_\text{F}(z,t)$ are real and negative, such that the full counting statistics $P_N$ can be expressed as a generalized binomial distribution with transmission probabilities $p_j=1/(1-z_j)$ that obey $p_j \in [0,1]$, such that the sign rule $(-1)^{m-1} C_{\text{F},m} \ge 0$ is fulfilled for all factorial cumulants.

We conclude this appendix by an additional remark that is based on the observation that, for a two-state system, the moment generating function Eq.~(\ref{eq:moment_generating_function_2_state}) can be expressed as a function of $\eta$ (or $\eta^2$), where the relation Eq.~(\ref{eq:eta}) between $\eta$ and $z$ implies $\eta^2=1$ for $z=1$ as well as $\partial_z (\eta^2) = 1-a^2$.
This immediately yields 
\begin{align}
    C_{\text{F},m}=(1-a^2)^m \left.\partial_{\eta^2}^m \ln \mathcal{M}_\text{F}(\eta^2)\right|_{\eta^2=1} \,.
\end{align}
For vanishing asymmetry ($a=0$), Eq.~(\ref{eq:eta}) implies $\eta^2=z$.
We, therefore, find the scaling law
\begin{align}
    \frac{C_{\text{F},m}}{(1-a^2)^m} = \left. C_{\text{F},m} \right|_{a=0} \, .
\end{align}
As a consequence, the validity of the sign rule for the \textit{symmetric} two-state model is equivalent to its validity for the \textit{asymmetric} case as well.

\section{Different representations of discrete probability distributions}
\label{sec: Representation of arbitrary discrete probability distributions}

In this appendix, we summarize how the different representations of a discrete probability distribution $P_N$ employed in this paper are connected to each other. 
These connections are established through the $z$-dependence of the moment generating function $\mathcal{M}_{\text{F}}(z)$ (if it exists) or, equivalently, the cumulant generating function $\mathcal{S}_{\text{F}}(z)= \ln \mathcal{M}_\text{F}(z)$.
In terms of $P_N$, the moment generating function reads
\begin{align}
    \mathcal{M}_\text{F}(z)=\sum_{N}z^NP_N 
\end{align}
with the normalization condition $\mathcal{M}_{\text{F}}(1)=1$.

For any distribution $P_N$ that is truncated at a finite $N_\text{max}$, the moment generating function $\mathcal{M}_{\text{F}}(z)$ becomes a polynomial of order $N_\text{max}$, which is fully characterized by its (in general complex-valued) zeros $z_j$,
\begin{align}
    \mathcal{M}_\text{F}(z)=\prod_j \left(
    \frac{z-z_j}{1-z_j} \right)  \, .
\end{align}
By defining $p_j:=1/(1-z_j)$, we achieve as a third representation 
\begin{align}
    \mathcal{M}_\text{F}(z)=\prod_j \left( 1-p_j+zp_j\right)  \, .
\end{align}
A fourth representation
\begin{align}
    \mathcal{M}_\text{F}(z)= \exp \left[ \sum_{m} C_{\text{F},m} \frac{(z-1)^m}{m!}  \right] 
\end{align}
is based on the factorial cumulants 
$C_{\text{F},m}=\left.\partial_z^m \ln \mathcal{M}_\text{F}(z)\right|_{z=1}$.
The equivalence of these representations can be used to express the factorial cumulants $C_{\text{F},m}$ in terms of the $p_j$'s, as done in Eq.~(\ref{eq:factorial cumulants generalized binomial}).

The representation of the moment generating function $\mathcal{M}_\text{F}(z)$ through its zeros $z_j$ (or, equivalently, $p_j$) is always possible for any finite but not necessarily for an infinite $N_\text{max}$.
An example is the Poisson distribution $P_N^{\text{\,Poisson}}= (\Gamma t)^N e^{-\Gamma t}/N!$, since its moment generating function $\mathcal{M}_\text{F}^{\text{\,Poisson}}(z)= \exp \left[ (z-1) \Gamma t \right]$ does not have any zero at all.
The Poisson distribution can be understood as the $N_\text{max}\to \infty$ limit of a binomial distribution for $N_\text{max}$ attempts with success probability $p=(\Gamma t)/N_\text{max}$.
The corresponding moment generating function has a single but $N_\text{max}$-fold zero at $z=1- \frac{N_\text{max}}{\Gamma t}$, which moves to $-\infty$ for $N_\text{max}\to \infty$.

For a generalized binomial distribution, $p_j$ plays the role of a probability, which implies $p_j \in [0,1]$, such that $z_j$ lies on the negative real axis.
Each set of given probabilities $p_j$ corresponds to a valid distribution $P_N$, for which there are more efficient ways to calculate it \cite{biscarriSimple2018} than simply applying Eq.~\eqref{eq:generalized binomial}.
In many situations, an explicit calculation of $P_N$ is not necessary, but a direct calculation of the factorial cumulants using Eq.~\eqref{eq:factorial cumulants generalized binomial} is advantageous.

Distributions $P_N$ that are not of generalized binomial type feature zeros $z_j$ of the moment generating function \textit{outside} the negative real axis, such that $p_j$ can no longer be interpreted as a probability.
Not each set of complex-valued zeros $z_j$ lead to a valid distribution $P_N$.
For example, it is clear that non-real zeros have to come in complex-conjugate pairs to guarantee that $P_N$ remains real.
Furthermore, if the moment generating function has only a single pair $z_\pm=x\pm iy$ of complex-conjugated zeros, their real parts must obey $x\le 0$ since, otherwise, the probability $P_1=-\frac{2x}{(1-x)^2+y^2}$ would become negative.

\section{Influence of the limited time resolution on the factorial cumulants}
\label{sec:Influence of the limited time resolution on the factorial cumulants}

\begin{figure}
    \centering
    \includegraphics[width=0.95\linewidth]{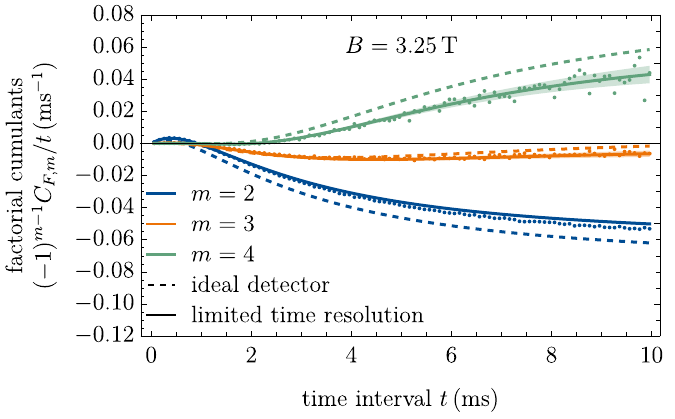}
    \caption{Comparison between the model for an ideal detector (dashed lines) and the model including the limited time resolution (straight lines) of $t_\text{bin}=60\,\mu$s.}
    \label{fig: time resolution}
\end{figure}

The master equation and its solution Eq.~\eqref{eq:generating function solution} model an experiment with an ideal detector. Errors caused by limited time resolution $t_\text{bin}$ can be modeled by discretizing the master equation  \cite{kleinherbersPushing2022,kleinherbersUnraveling2023,mannelQuantum2025}. This leads to a new solution
\begin{align}
    \mathcal{S}_{\text{F}}^{(t_\text{bin})}(z,t)=\ln\left(\boldsymbol{e}^T\boldsymbol{\Pi}_z^{t/t_\text{bin}}\boldsymbol{P}_\text{st}\right)
    \label{eq:generating function solution tbin}
\end{align}
including either the propagator
\begin{align}
    \boldsymbol{\Pi}_z=\begin{pmatrix}
        1&&z&&z\\
        1&&1&&1\\
        1&&1&&1
    \end{pmatrix}
    \circ e^{\boldsymbol{W}t_\text{bin}}
\end{align}
for the three-state model or the propagator
\begin{align}
    \boldsymbol{\Pi}_z=\begin{pmatrix}
        1&&z\\
        1&&1
    \end{pmatrix}
    \circ e^{\boldsymbol{W}t_\text{bin}}.
\end{align}
for the two-state model. The symbol $\circ$ denotes the element-wise multiplication of two matrices. The comparison between the model for the ideal detector from Eq.~\eqref{eq:generating function solution} (dashed lines) and the model including the limited time resolution (straight lines) following Eq.~\eqref{eq:generating function solution tbin} is shown for $B=3.25\,$T in Fig.~\ref{fig: time resolution}.
We find that the missed events due to the finite time resolution leads to small quantitative but not to any qualitative changes of the factorial cumulants.
In particular, the violation of the sign rule is not affected.

As another imperfection, the detector is intrinsically noisy since the distribution of binned photon counts has a nonvanishing width.
This may result in falsely identified events \cite{kleinherbersPushing2022}. 
However, we chose the binning time  $t_\text{bin}$ large enough to make the detector noise negligible.

\section{Factorial versus ordinary cumulants}

In our study, we use \textit{factorial} cumulants in order to extract information from the probability distribution $P_N$. 
They can be expressed in terms of \textit{ordinary} moments $\left\langle N^m \right\rangle=\sum_{N}N^mP_N$ of the distribution.
Using Eq.~\eqref{eq:generating function}, the first four factorial cumulants can be written as
\begin{subequations}
\begin{align}
    C_{\text{F},1} = &  \langle N \rangle,
    \\
    C_{\text{F},2} = & \langle N^2 \rangle - \langle N \rangle^2 - \langle N \rangle,
    \\
    \begin{split}
    C_{\text{F},3} = & \langle N^3 \rangle - 3\langle N^2 \rangle \langle N \rangle + 2 \langle N \rangle^3 \\&- 3 \langle N^2 \rangle + 3 \langle N \rangle^2 + 2 \langle N \rangle,
    \end{split}\\
    \begin{split}
    C_{\text{F},4} = & \langle N^4 \rangle-4 \langle N^3 \rangle \langle N \rangle-3 \langle N^2 \rangle^2 \\& +12 \langle N \rangle^2 \langle N^2 \rangle -6 \langle N \rangle^4-6 \langle N^3 \rangle \\& +18
   \langle N \rangle \langle N^2 \rangle-12 \langle N \rangle^3 +11
   \langle N^2 \rangle \\& -11 \langle N \rangle^2-6 \langle N \rangle.
    \end{split}
\end{align}
\end{subequations}

Probability distributions are frequently characterized by \textit{ordinary} cumulants.
They describe deviations from a Gauss distribution and are, therefore, well suited for continuous rather than discrete stochastic variables.
The first four ordinary cumulants are given by the expectation value $\kappa_1=\langle N \rangle$, the variance $\kappa_2=\langle N^2 \rangle-\langle N \rangle^2$, the skewness ${\kappa_3=\langle N^3 \rangle-3\langle N^2 \rangle\langle N \rangle+2\langle N\rangle^3}$, and $\kappa_4= \langle N^4 \rangle-3\langle N^2 \rangle^2-4\langle N \rangle \langle N^3 \rangle+12 \langle N \rangle^2 \langle N^2 \rangle-6 \langle N \rangle^4$, which is related to the kurtosis. 
Ordinary cumulants are less suited than the factorial ones to check whether or not an effective single-particle picture is applicable.
One reason is that they do not obey a sign rule for generalized binomial distributions. 
In contrast, ordinary cumulants are plagued by universal oscillations \cite{flindtUniversal2009} with frequent sign changes. 
In addition, factorial cumulants have proven to be more resilient to errors caused by detector imperfections \cite{kleinherbersPushing2022}.

\bibliography{bibliography}

\end{document}